\documentclass[prb,twocolumn,preprintnumbers,amsmath,amssymb]{revtex4-1}
\usepackage{graphicx}

\begin{document}
\title{ Goldstone and Higgs modes of photons inside a cavity  }
\author{ Yi-Xiang Yu$^{1,2}$, Jinwu Ye $^{1,3}$  and Wuming Liu $^{2}$  }
\affiliation{
$^{1}$ Department of Physics and Astronomy, Mississippi State University, MS, 39762, USA  \\
$^{2}$ Institute of Physics, Chinese Academy of Sciences, Beijing, 100080, China   \\
$^{3}$ Department of Physics, Capital Normal University, Beijing, 100048 China  }

\date{\today }

\begin{abstract}
  Goldstone and Higgs modes have been detected in
  various condensed matter, cold atom and particle physics experiments.
  Here, we demonstrate that the two modes can also be observed in optical systems
  with only a few (artificial ) atoms  inside a cavity. We establish this connection by studying the $ U(1)/Z_2 $ Dicke
  model where  $ N $ qubits ( atoms ) coupled to a single photon mode.
  We determine the Goldstone and Higgs modes inside the super-radiant phase and their corresponding spectral weights
  by performing both $ 1/J=2/N $ expansion and exact diagonization (ED) study at a finite $ N $.
  We find nearly perfect agreements between the results achieved by the two approaches when $ N $ gets down even to $ N = 2 $.
  The quantum finite size effects at a few qubits make the two modes quite robust against an effectively small counter-rotating wave term.
  We present a few schemes to reduce the critical coupling strength, so the two modes can be
  observed in several experimental systems of (artificial ) atoms inside a cavity  by just conventional optical measurements.
\end{abstract}
\maketitle

{\sl Introduction: }
  It was well known that a broken global continuous symmetry in quantum phases \cite{anderson,sachdev}
  leads to two associated collective modes:
  the massless Goldstone mode and a massive Anderson-Higgs amplitude mode \cite{anderson,higgs} ( For topological ordered phases, see \cite{topo} ).
  The Goldstone modes  have been detected in a quantum anti-ferromagnet \cite{higgsmag}, a superfluid \cite{jinwusuper}
  and also in cold atom systems \cite{braggbog}.
  However, the massive Higgs mode and its decay rate are much more difficult to detect in experiments.
  Even so, the Higgs amplitude mode was detected in superconductors \cite{higgscdw,higgssuper} and in a quantum anti-ferromagnet \cite{higgsmag,higgssoft} near its quantum phase transition to a valence bond solid \cite{higgsmagvbs}.
  Unfortunately, due to the Galilean invariance in a superfluid,
  the phase mode and amplitude mode are conjugate variables, the conjugate pair only leads to a Goldstone mode, so
  there is no Higgs mode inside a superfluid \cite{higgssuper,jinwusuper}.
  Most recently, the Higgs amplitude mode and its decay rate were  detected in cold atoms loaded in  2 dimensional optical lattice
  near the superfluid to Mott transition \cite{higgslattice2d}.
  In a relativistic quantum field theory, it is  the well known Higgs mechanism \cite{higgs} which generates various mass spectrum of elementary particles.
  Although the various elementary particles have been discovered  with the predicted masses,
  the original massive Higgs particle stays elusive until it was tentatively discovered with its mass $ \sim 125 GeV $ and width
  $ \sim 6 MeV $ in the recent LHC experiments \cite{higgslhc}.

\begin{figure}
\includegraphics[width=8.5cm]{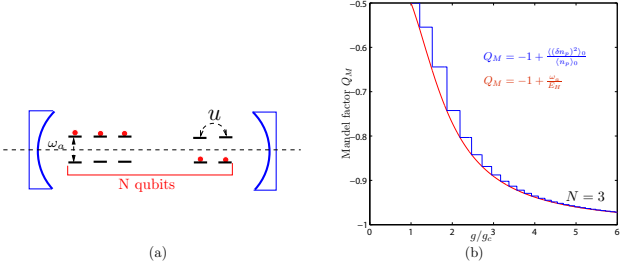}
\caption{
(a) $N$ (artificial) atoms are placed on anti-nodes of a cavity. $ u $ is the repulsive qubit-qubit interaction
which can be tuned to reduce the critical coupling $ g_c $.
(b) The analytical Mandel factor $Q_M$ (red) against the ED result (blue) at $ N=3 $. It is a number squeezed state inside the superradiant phase.}
\label{mandel}
\end{figure}

  In this paper, we will present the first study of the Goldstone and Higgs modes of photons inside a cavity.
  The system is described by the $ U(1)/Z_2 $ Dicke
  model Eqn.\ref{u1} where  $ N $ cold atoms \cite{orbitalt,orbital,fewboson,fewfermion}, qubits \cite{qubitweak,qubitstrong} and quantum dots \cite{dots}
  coupled to a single photon mode inside a cavity (Fig.\ref{mandel}a).
  It was known that in the thermodynamic limit \cite{dicke1,popov,bethe,staircase,berryphase,chaos,gprime1,gprime2}, when
  the atom-photon coupling $ g $ is sufficiently large,
  the system undergoes a quantum phase transition from a normal phase to a emergent superradiant phase which breaks the global
  ( or approximate ) $ U(1)$ symmetry.
  We perform both $ 1/J=2/N $ expansion and exact diagonization (ED)
  study on how  the Goldstone mode and Higgs amplitude mode inside the superradiant phase evolves as the $ N $ decreases to a few.
  We find nearly perfect agreements between the results achieved from the $ 1/J $ calculations
  with those from the ED studies in all physical quantities even when $ N $ gets down even to $ N = 2 $.
  The system's energy levels in the super-radiant phase display a Landau-level like structure
  with the inter-Landau energy scale setting by the Higgs energy $ E_{H} $ and
  the intra-Landau energy scale setting by the Goldstone energy $ E_G $.
  In both the photon and photon number correlation functions, we evaluate the low frequency Goldstone mode $ E_G $,
  the high frequency Higgs mode $ E_H $ and their  corresponding spectral weights $ C_G $ and $ C_H $.
  The Higgs mode is a sharp mode protected by the $ U(1) $ symmetry at any finite $ N $.
  We also study the effects of the counter rotating wave ( CRW ) term by the $ 1/J $ expansion and find that
  the quantum finite size effects at a few qubits $ N \sim 2-5 $ make the two modes robust against the CRW  term if $ g^{\prime}/g < 1/3 $.
  We  discuss several schemes to reduce the critical coupling considerably, so the two modes can be observed in several
  experimental systems by conventional optical detection methods such as the florescence spectrum measurement [58] on Eqn.\ref{aacorr} and
  the  HanburyBrown-Twiss (HBT) type of measurement [59] on Eqn.\ref{nncorr} respectively.

{\sl Reducing the $ U(1)/Z_{2} $ to the $ J-U(1)/Z_{2} $ Dicke model:}
  In the $ U(1)/Z_2 $ Dicke model \cite{dicke}, a single mode of photons couple to $
  N $ two level atoms with same coupling  constants $ \tilde{g} $ and $ \tilde{g}^{\prime} $. The two level
  atoms can be expressed in terms of 3 Pauli matrices $ \sigma_{\alpha},
  \alpha=1,2,3 $.  The $ U(1)/Z_{2} $ Dicke model can be written as:
\begin{eqnarray}
  H_{U(1)/Z_2} & = & \omega_{a} a^{\dagger} a +  \frac{\omega_{b}}{2} \sum^{N}_{i=1} \sigma^{z}_{i}
   + \frac{g}{\sqrt{N}}\sum^{N}_{i=1} ( a^{\dagger} \sigma^{-}_{i} + h.c. )   \nonumber  \\
   & + & \frac{g^{\prime}}{\sqrt{N}}\sum^{N}_{i=1} ( a^{\dagger} \sigma^{+}_{i} + h.c. )
\label{u1}
\end{eqnarray}
   where the $ \omega_a, \omega_b $ are the cavity photon frequency and the energy difference of the two atomic levels  respectively,
   the  $  g= \sqrt{N} \tilde{g} $ is the collective photon-atom coupling (  $  \tilde{g} $ is the individual photon-atom coupling ).
   The $  g^{\prime}= \sqrt{N} \tilde{g}^{\prime} $ is the counter-rotating wave term.
   It was demonstrated in \cite{gprime1,gprime2} that in the thermal or cold atom experiments \cite{orbitalt,orbital},
   the strengths of $ g $ and $ g^{\prime} $ can be tuned separately by using circularly polarized pump beams in  a ring cavity.
   In the qubit \cite{qubitweak,qubitstrong} or quantum dot \cite{dots} experiments,
   the CRW terms and RW terms have the same strength at the bare level, however,
   the CRW term is usually much smaller than the RW term at the effective level as is the case in the experiment \cite{qubitweak}.
   This is because the former violates the energy conservation, while the latter respects the energy conservation.
   However, when the coupling strength gets close to the the transition frequency,
   the CRW term becomes comparable to the RW term as is the case in the experiment in \cite{qubitstrong}.
   In any case, the Hamiltonian Eqn.\ref{u1} with independent $ g $ and $ g^{\prime} $ is the most general
   Hamiltonian describing various experimental systems in various coupling regimes under the two atomic levels and a single photon mode approximation.


   One can introduce the total "spin"  of the $ N $ two level
   atoms $ J^{z}= \sum_{i} \sigma^{z}_{i}, J^{+}= \sum_{i}  \sigma^{+}_{i}, J^{-}= \sum_{i} \sigma^{-}_{i} $,
   When all the $ N $  atoms are in the ground state,
   then $ J= N/2, J_{z}=-N/2 $, because the total spin $ J^{2}= J^{2}_{x}+
   J^{2}_{y}+J^{2}_{z} $ is a conserved quantity,  by
   confining the Hilbert space only to $ J=N/2 $, then one reduces the Hilbert space from $ 2^{N} $ to $ 2J+1=N +1 $.
   One can call the resulting model as the $ J-U(1)/Z_2 $ Dicke model.
   One main advantage of this reduction is that one can
   study the  $ J-U(1)/Z_2 $ model by using Holstein-Primakoff (HP) representation of the angular momentum operator  $ J_{z}=
   b^{\dagger}b-J, J_{+}= b^{\dagger} \sqrt{ 2 J- b^{\dagger} b}, J_{-}=  \sqrt{ 2 J- b^{\dagger} b} b $, therefore
   treat photon and atom on the same footings.
   This advantage will enable us to bring out many new and important results
   hard to retrieve  from the $ 1/N $ expansion in \cite{berryphase}. Very fortunately, this reduction will not change the most important physics
   of the original $ U(1)/Z_2 $ model Eqn.\ref{u1}.
   As argued in  Supplementary materials B and explicitly shown in \cite{u1nature}, except the $ U(1)/Z_2 $ Dicke model contains some additional
   energy levels, both models share the same other physical quantities to be studied in this paper.

   If $ g^{\prime}=0 $,  the Hamiltonian Eqn.\ref{u1} has the $ U(1) $ symmetry
   $ a \rightarrow  a  e^{ i \theta}, \sigma^{-} \rightarrow \sigma^{-} e^{ i \theta} $.
   The CRW $ g^{\prime} $ term breaks the $ U(1) $ to the $ Z_2 $ symmetry $ a \rightarrow -a , \sigma^{-} \rightarrow -\sigma^{-} $.
   If $ g^{\prime} =g  $, it become the $ Z_2 $ Dicke model studied in \cite{chaos}.
   In this paper, we focus on the $ U(1) $ Dicke model, but will also consider the effects of the small counter-rotating wave term $ g^{\prime} < g $
   in the experimental detection section and the supplementary materials Sec.D.
   The $ g^{\prime}=g $ and the $  g^{\prime} \sim g $ cases will be studied in \cite{chaos}.
   The $ U(1) $ Dicke model was solved in the thermodynamic limit $ N=\infty $ by various methods \cite{dicke1,popov,bethe,staircase,berryphase,chaos}.
   In the normal phase $ g < g_c =\sqrt{\omega_a\omega_b } $, $ \langle a \rangle =0 $, the $
   U(1) $ symmetry is respected. In the super-radiant phase $ g > g_c $,
   $ \langle a \rangle \neq 0 $, the $ U(1) $ symmetry is spontaneously broken.

{\sl  Goldstone and Higgs modes  in the super-radiant phase by 1/J expansion: }
   In the super-radiant phase $ g > g_c $ and also not too close to the quantum critical point (QCP) ( if
   too close, then $ a^{\dagger} a \ll j, b^{\dagger} b \ll j $, a direct $ 1/j $ expansion is needed and will be performed elsewhere ),
   it is convenient to write
   both the photon and atom in the polar coordinates $ a= \sqrt{
   \lambda^{2}_{a} + \delta \rho_a } e^{ i \theta_a}, b= \sqrt{
   \lambda^{2}_{b} + \delta \rho_b } e^{ i \theta_b} $ where $ \lambda^{2}_{a} \sim \lambda^{2}_{b} \sim j $. When
   performing the controlled $ 1/J $ expansion, we keep the terms to the order
   of $ \sim j, \sim 1 $ and $ \sim 1/j $, but ignore orders of $ 1/j^{2} $.
   We first minimize the ground state energy at the order $ j $, we
   found the saddle point values of $ \lambda_a $ and $ \lambda_b $:
   $ \lambda_a= \frac{g}{ \omega_a } \sqrt{ \frac{j}{2} ( 1 - \mu^{2} ) },~~~~ \lambda_b= \sqrt{ j(1-\mu) }
   $ where $  \mu =g^{2}_c/g^{2}= \omega_a \omega_b/g^{2} $. It
   holds only in the superradiant phase $ g > g_c $.

\begin{figure}
\includegraphics[width=6cm]{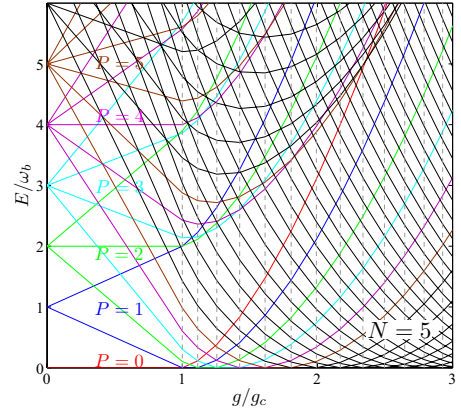}
\caption{ The ED results (See the Methods section) of the energy levels $E$ measured by subtracting the ground-state energy versus $g/g_c$ at
 resonance $\omega_a=\omega_b$ with $ N=5 $ atoms.
 Different colors of the energy curves correspond to several smallest numbers of total excitations number
 $ P=a^{\dagger} a + b^{\dagger} b $. The dashed vertical lines correspond to the critical values of $g$ where the number of
 total excitations $ P $ in the ground state increases by one.}
\label{spectrum}
\end{figure}

   Observe that (1) in the superradiant phase $ g > g_c $, $ \lambda^{2}_a \sim
   \lambda^{2}_b \sim j $, (2) it is  convenient to get to the $ \pm
   $ modes:
   $ \theta_{\pm}= (\theta_a \pm \theta_b)/2, \delta \rho_{\pm}= \delta \rho_a \pm \delta \rho_b,
   \lambda^{2}_{\pm}= \lambda^{2}_a \pm  \lambda^{2}_b $.
   (3) paying a special attention to the crucial Berry phase term in the $ \theta_{+} $ sector,
   (4) after shifting $ \theta_{-} \rightarrow \theta_{-} + \pi/2 $,  then one can
   get the effective action up to the order of $ 1/j $:
\begin{eqnarray}
 {\cal L}_{ U(1) }[ \delta \rho_{\pm}, \theta_{\pm} ]  =  i ( \lambda^{2}_{+} + \delta
  \rho_{+} ) \partial_{\tau} \theta_{+} + i ( \lambda^{2}_{-} + \delta
  \rho_{-} ) \partial_{\tau} \theta_{-}    \nonumber  \\
   + \frac{ D }{2} (\delta \rho_{+} )^{2} + D_{-} (\delta \rho_{-} + \gamma \delta \rho_{+} )^{2}
   +  4 \omega_{a} \lambda^{2}_a \sin^{2} \theta_{-} ~~~~~
\label{pmu10}
\end{eqnarray}
   where the first line are the crucial Berry term in the $ \theta_+ $ and $ \theta_- $ respectively,
   $ D=  \frac{ 2 \omega_a g^{2} }{  E^{2}_{H} N }  $ is the phase diffusion constant,
   $ D_{-}= E^{2}_{H}/16 \lambda^{2}_{a} \omega_a $ with $ E^{2}_{H}= ( \omega_a+\omega_b)^{2} + 4 g^2 \lambda^{2}_{a}/N $.
   The $ \gamma= \frac{ \omega^{2}_{a} }{ E^{2}_{H} } ( 1- \frac{g^{4}}{ \omega^{4}_{a} } ) $ is the coupling between the
   $ + $ and $ - $ sector.  Under the $ U(1) $ transformation $ \theta_{a/b} \rightarrow \theta_{a/b} + \chi $,
   $ \theta_{+} \rightarrow  \theta_{+} +  \chi, \theta_{-} \rightarrow  \theta_{-} $, so
   the $ \theta_{-} $ is neutral under the $ U(1) $ transformation.
   There is a mass term for $ \theta_{-} $, but no mass term for $ \theta_{+} $.
   The conjugate pair $ ( \theta_{+}, \delta \rho_{+} ) $ leads to the Goldstone mode $ E_{G} $ as shown in Eqn.\ref{aacorr}.
   While  the conjugate pair $ (\theta_{-}, \delta \rho_{-} ) $ leads to the Higgs mode $ E_{H} $ as shown in Eqn.\ref{nncorr}
   ( See also Supplementary materials C ).
\begin{figure}
\includegraphics[width=8.5cm]{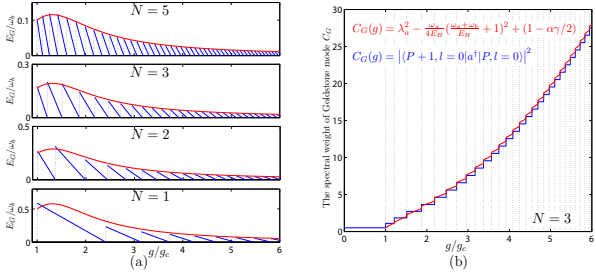}
\caption{(a) The analytical Goldstone mode at $ \alpha =-1/2 $, $ E_{G}(\alpha =-1/2)= D(g)= \frac{ 2 \omega_a g^{2} }{  E^{2}_{H} N }$ (red line)
 are contrasted with the ED result $ E_G= E^{P+1}_{0}- E^{P}_{0} $ (blue lines)  at $ N=5,3,2,1$ respectively.
 It is remarkable that the analytical result can even map out broad peaks at small $ P $ in the ED results very precisely.
 (b) The analytical spectral weight (red) of the Goldstone mode $ C_G $ against the ED result (blue) at $ N=3 $.}
\label{goldstone}
\end{figure}

   Defining the Berry phase in the $ + $ sector as $ \lambda^{2}_{+} = P + \alpha $ where $ P=1,2,\cdots $ is the closest integer  to the $ \lambda^{2}_{+} $, so $ -1/2 < \alpha < 1/2 $. In fact, $ P=a^{\dagger} a + b^{\dagger} b $ is just the
   conserved total excitations number. Redefine $  \delta \rho_{+}= \hat{N}- P $, then one can write the corresponding
   Hamiltonian of Eqn.\ref{pmu10} as:
\begin{equation}
    H_{U(1)}= \frac{ D }{2} (\delta \rho_{+} - \alpha )^{2}
     + D_{-} [\delta \rho_{-} + \gamma  \delta \rho_{+} ]^{2}
    +  4 \omega_{a} \lambda^{2}_a \sin^{2} \theta_{-}
\label{u1h}
\end{equation}

Because the $ \theta_{-} $ is very massive, after pinning $ \theta_{-} $ around $ \theta_{-} \sim 0 $,
one can approximate $ \sin^{2} \theta_{-} \sim  \theta^{2}_{-} $,
     so the total wavefunction is $
      \psi_{l,m}( \theta_{+},\theta_{-} )=\frac{1}{ \sqrt{2 \pi}} e^{ i [ ( m + l ) \theta_{+} + \gamma( m + l ) \theta_{-} ] } \psi_{l} ( \theta_{-} )
     $  where  the $ l=0,1,\cdots $ are the Landau level indices, the $ m=-P, -P+1, \cdots $ are the magnetic indices at a given sector $ P $,
     $ 0 < \theta_{+} < 2 \pi, -\infty < \theta_{-} < \infty $ and the
     $ \psi_{l} ( \theta_{-} ) $ is just the $ l $-th the wavefunction of a harmonic oscillator.

     The corresponding eigen-energy is
\begin{equation}
     E_{0} ( l, m) = ( l+ 1/2 ) \hbar E_H + \frac{D}{2} ( m + l -\alpha )^{2}
\label{u1energy0}
\end{equation}
    The ground state energy is at $ l=0, m=0 $.

    One can see that the energy spectrum Eqn.\ref{u1energy0} has a Landau-level structure:
    the Landau level energy scale is given by
    the Higgs energy $ E_{H} \sim 1 $, the intra-Landau level is set up by the Goldstone energy scale $ E_{G} \sim 1/j $.
    In the large $ j $ limit, there is a wide separation of the two energy scales $ E_H \sim 1 \gg E_G \sim 1/j $.
    When the excitation number $ P $ reaches the order of $ N $, then the intra-Landau levels with $ |m| \ge P $ will
    start to overlap with the inter-Landau levels. These analytical results explain precisely the
    ED energy level structures  shown in Fig.\ref{spectrum} for the resonant case $ \omega_a=\omega_b $.

    Away from the QCP, one can write down the $ 1/j $ expansion of the atom operator:
   $ a  =   [ \lambda_a   + \frac{ \delta \rho_a }{ 2 \lambda_a } -   \frac{ (\delta \rho_a)^{2} }{ 8 \lambda^{3}_a } + \cdots ] e^{ i \theta_a } $.
   At a finite $ N $, due to the restoration of the $ U(1) $ symmetry by the phase diffusion in the $ \theta_{+} $ sector,
   any $ U(1) $  non-invariant correlation functions vanish $
      \langle  a \rangle =0,~~~ \langle  a(\tau) a (0) \rangle  =0 $
    So we need only focus on the $ U(1) $ invariant correlation functions.
    By using both canonical quantization and path integral approaches,  we find the single photon correlation function \cite{u1nature}:
\begin{eqnarray}
    \langle {\cal T} a(\tau) a^{\dagger} (0) \rangle   =   C_{G} e^{-E_{G} \tau} +  C_o e^{-E_o \tau} + O(1/j)     \nonumber   \\
    C_{G}  =  \lambda^{2}_a - C_o + ( 1- \gamma \alpha/2 ) ,~~E_{G}=D (\frac{1}{2}- \alpha)     \nonumber   \\
    C_o  =  \frac{ \omega_a }{ 4 E_{H} } ( \frac{ \omega_a + \omega_b }{  E_{H} } + 1 )^{2},~~E_o= E_{H} + E_{G}
\label{aacorr}
\end{eqnarray}
    where $ E_G= D (\frac{1}{2}- \alpha) $ is the Goldstone mode with the corresponding spectral weight $ C_G $, while
    $ E_o= E_H + E_G $ is the optical mode with the corresponding spectral weight $ C_o $.
    All these quantities can be directly measured by the florescence spectrum measurement [58].

\begin{figure}
\includegraphics[width=8.5cm]{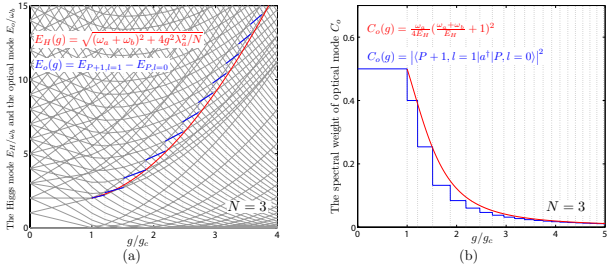}
\caption{(a) The analytical relation $ E_o= E_{H} + E_G $ ($ E_H $ in red line) is satisfied
 by the ED optical mode $ E_o =E^{P+1}_{1}-E^P_0 $ (blue lines) at $ N=3 $ except at the first few steps.
(b) The analytical spectral weight (red) of the optical mode $ C_o $ against the ED result (blue) at $N=3$.}
\label{optical}
\end{figure}

  The $ E_G, C_G $ and $ E_o, C_o $ are compared with the ED results in Fig.\ref{goldstone} and Fig.\ref{optical} respectively.
  One can see that except at the first few $ P \ll N $ steps, the ED in $ E_o $ match the analytical relation $ E_o=E_{H}+E_{G} $
  in Eqn.\ref{aacorr} well. The discrepancy at the first few steps is not surprising, as said previously, if too close to the QCP,
  a direct $ 1/j $ expansion is needed and will be performed elsewhere.
  However, the agreement  between the analytical and ED results in $ C_o $ holds in all couplings even near the QCP.


\begin{figure}
\includegraphics[width=8.5cm]{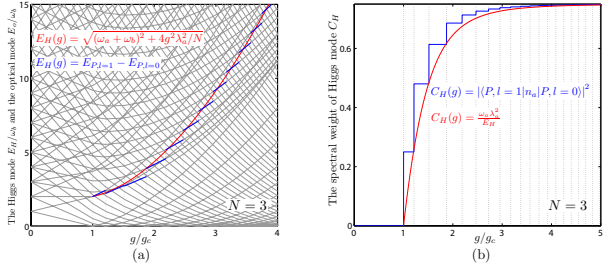}
\caption{(a) The analytical Higgs energy $ E_{H} $ (red) against the ED result $ E_{H}=E^{P}_{1}-E^P_0 $ (blue) at $ N=3 $.
(b) The analytical spectral spectral weight  $ C_{H} $ (red) for the Higgs mode
 against the ED result (blue) at $ N=3 $.}
\label{higgs}
\end{figure}

     One can also compute the photon number correlation function:
\begin{equation}
\langle {\cal T} n_{a}(\tau) n_a (0) \rangle - \langle n_a \rangle^{2}   =
\langle \delta \rho_a(\tau) \delta \rho_a(0) \rangle  =   \frac{ \omega_a \lambda^{2}_{a} }{ E_{H} } e^{- E_{H} \tau }
\label{nncorr}
\end{equation}
     where $ \langle n_a \rangle=\lambda^{2}_a $.
     The Higgs energy $ E_{H} $  and the corresponding spectral weight
     $ C_{H}=  \frac{ \omega_a \lambda^{2}_{a} }{ E_{H} } $ are compared with the ED results in Fig.\ref{higgs}.
     Note that the sharpness of the Higgs mode is protected by the conservation of $ \delta \rho_{+} $ in Eqn.\ref{u1h}.
     Both $ C_H $ and $ E_{H} $ can be directly measured by the  HanburyBrown-Twiss (HBT) type of measurement on two photon correlation functions [59].

    From the  Eqn.\ref{nncorr}, one can see that
    $  \langle (\delta \rho_a )^{2} \rangle  = \frac{ \omega_a \lambda^{2}_{a} }{ E_{H} } $, so one can find the Mandel $ Q $ factor:
   $ Q_{M}= -1+ \frac{ \omega_a }{E_{H}} $ which was compared with the ED result in the Fig.\ref{mandel}b.
     For $ \omega_a=\omega_b $, one can see $ -1 < Q_M < -1/2 $. So it is always in a number squeezed state.
     As $ g \rightarrow \infty $ limit, $ Q_M \rightarrow -1 $, so it approaches a photon Fock state.
     It is known that number squeezed states could be very important in quantum information processing and also
     in high-resolution and high sensitivity measurements.   Very similarly, one can evaluate the  atom correlation functions.


{\sl Effects of the CRW term and experimental detections of the Goldstone and Higgs modes: }

  The effects of the CRW terms on system's energy Eqn.\ref{u1energy0}, photon correlation function Eqn.\ref{aacorr} and the number correlation function Eqn.\ref{nncorr} are examined in the supplementary materials Sec.D. Their effects were found to be much smaller than those of the finite
  size for a few qubits $ N \sim 2-5 $ if $ g^{\prime}/g < 1/3 $.
  Recent experiments \cite{orbitalt,orbital} reached the $ Z_2 $ super-radiant regime \cite{chaos} with the help of a transverse pumping.
  In this transverse pumping scheme, the CRW terms in Eqn.\ref{u1} are as important as the RW ones $ g^{\prime}=g $, so only the $ Z_2$ super-radiant phase
  can be realized. However, it was demonstrated in \cite{gprime1,gprime2} that the strengths of $ g^{\prime} $ and $ g $
  can be tuned independently by using circularly polarized pump beams in a ring cavity. So we expect that $ g^{\prime}/g < 1/3 $
  can be achieved in this transverse pumping scheme, then the system can be tuned to the $ U(1) $ superradiant regime.
  It is also promising to reach the $ Z_2 $ super-radiant regime "simultaneously"
  ( namely without any transverse pumping ) with artificial atoms such as
  superconducting qubits inside micro-wave circuit cavity \cite{qubitweak,qubitstrong} and  quantum dots inside a
  semi-conductor nano-cavity engraved in a photonic crystal in Fig.\ref{mandel}a \cite{dots}.
  Indeed, very recently, by enhancing the inductive coupling of a flux qubit to a transmission line resonator,
  a remarkable ultra-strong  coupling with individual $ \tilde{g} \sim 0.12 \omega_a $ was realized in a circuit QED system \cite{qubitstrong}.
  In this simultaneous scheme, due to the violation of the energy conservation, the CRW term is usually much smaller
  than the RW one $ g^{\prime}< g $, but gets stronger as the coupling gets stronger.
  In real experiments of superconducting qubits or quantum dots inside a cavity in Fig.\ref{mandel}a, there are always the potential
  scattering term $  \lambda_z  J_{z} a^{\dagger} a/j $  between the cavity photons and the qubits and the qubit-qubit
  interaction term $ u J^{2}_{z}/j $. The critical coupling  $g_c$ is shifted to:
\begin{equation}
  g_c + g^{\prime}_c = \sqrt{(\omega_a-\lambda_z)(\omega_b-2u)}
\label{uugc}
\end{equation}
  which indicates that the two repulsive interaction terms can be used to decrease the critical $ g_c$ well below the
  bare critical strength $ \sqrt{ \omega_a \omega_b} $. The qubit-qubit interactions can be tuned inductively or capacitively.
  This fact could be used to put the system into the regime where the CRW term satisfies $  g^{\prime}/g < 1/3 $,
  so the $ U(1) $ super-radiant phase can be realized in the possible future experiments using both atoms inside a optical cavity
  or qubits inside a microwave circuit QED in Fig.\ref{mandel}a.

{\sl Conclusions:}
  Quantum mechanics describes the motion of a single or a few particles \cite{nobel1,nobel2}.
  Condensed matter physics studies various emergent quantum phenomena of macroscopic number of interacting particles. Ultracold atom systems and optical cavity systems can provide unprecedented experimental systems to
  study quantum phenomena ranging from a few  particles to a million number of interacting particles.
  Due to the tremendous tunability of all the parameters
  in these systems, they can be tuned to scale up from the isolated quantum mechanics systems to macroscopic condensed matter systems.
  The conventional route is to look at how " more is different " emerges, namely, study
  how various macroscopic quantum phenomena emerge as the number of particles gets " more and more "\cite{anderson}.
  Here, we have taken a dual point of view: study how the emergent phenomena evolve as the number of particles becomes " less and less ".
  This dual approach becomes especially important in view of recent experiments of  cold atoms inside an optical cavity \cite{orbitalt,orbital}
  or superconducting qubits \cite{qubitweak,qubitstrong} or quantum dots \cite{dots} inside a microcavity, involving only finite to
  even small number of particles (Fig.\ref{mandel}a), also manipulating only a few atoms in current experiments \cite{fewboson,fewfermion}. Specifically,
  we studied how the emergent Goldstone and Higgs modes evolve as the number of particles gets less and less, even down only a few particles
  in  quantum optical systems.  In general, many body theory developed to study
  the emergent phenomena of condensed matter systems can also be a very powerful tool to study the physical phenomena from millions
  of particles down even to a few particles. Our theoretical works should provide a solid foundation for
  various ongoing and upcoming systems with a small number of particles to observe the novel phenomena due to strong light-matter interactions
  explored in this report.

{\bf Acknowledgements:}
We thank Yu Chen for his participation  in the early stage of the project and his contributions leading to Eqn.2.
Prof. Guangshan Tian for encouragements. JY thanks D. Podolsky, Han Pu and Jiangqiang You for helpful discussions.
Y.Y and JY are supported by NSF-DMR-1161497, NSFC-11074173, -11174210,
Beijing Municipal Commission of Education under Grant No.PHR201107121, at KITP is supported in part by the NSF under grant No. PHY11-25915.
W.M.Liu is supported by NSFC under Grants No. 10934010 and No. 60978019, the NKBRSFC under Grants No. 2012CB821300.

{\bf Supplementary materials }

{\sl A. Exact Diagonalization ( ED) study: }

  For simplicity,  in the following, we limit our ED study only to the resonant case $ \omega_a=\omega_b $.
  We assume $ P \leq N $. The $ P > N $ case can be similarly addressed by changing $ P+1 $ to $ N +1 $.
  The ground state in the given $ P $ Hilbert space is:
\begin{equation}
   | P, G \rangle=| P, l=0 \rangle = \sum^{P}_{s=0}  A^{P,l=0}_{s} | N/2, s-N/2  \rangle_{A} | P-s  \rangle_{F}
\label{groundj}
\end{equation}
   where the coefficients $ A^{P,l=0}_{s} $ can be determined by the ED.
   From Eqn.\ref{groundj}, one can evaluate the Mandel $ Q $ factor $
   Q_{M}= -1+\langle (\delta n_{p})^{2} \rangle/\langle n_{p} \rangle $
   which was compared with the analytical result in Fig.\ref{mandel}b.

    The $ l$-th eigen-state in the $ P+1 $ sector with the eigen-energy $ E^{P+1}_{l}, l=0,1,\cdots,P+1$ is:
\begin{equation}
   | P+1, l \rangle= \sum^{P+1}_{s=0}  A^{P+1,l}_{s} | N/2, s-N/2  \rangle_{A} | P+1-s  \rangle_{F},~~~
\label{excitedj}
\end{equation}
   where the coefficients $ A^{P+1,l}_{s} $ can be determined by the ED.

    In the Lehmann representation, we can evaluate the photon-photon correlation function Eqn.\ref{aacorr}:
\begin{eqnarray}
    \langle {\cal T} a(\tau) a^{\dagger} (0) \rangle =  \sum^{P+1}_{l=0} e^{-(E^{P+1}_{l}-E^P_0)\tau} | \langle P+1, l|a^{\dagger} |P,G \rangle |^{2}
                                       \nonumber    \\
     \langle P+1, l|a^{\dagger} |P,G \rangle =  \sum^{P}_{s=0}  A^{*P+1,l}_{s} A^{P,0}_{s} \sqrt{ P+1-s}~~~~~~~
\label{aacorrnu}
\end{eqnarray}
    where $ E_{G}=E^{P+1}_{0}-E^P_0 $ is the Goldstone mode with the corresponding spectral weight
    $ C_{G}= | \langle P+1, l=0|a^{\dagger} |P, l=0 \rangle |^{2} $ , while
    $ E_o= E^{P+1}_{1}-E^P_0 $ is the optical mode with the corresponding spectral weight
    $ C_o=| \langle P+1, l=1|a^{\dagger} |P,l=0 \rangle |^{2} $ and so on.
    In fact, there are $ P+2 $ lines, we just focus on the two lowest energy excitations $ l=0,1$.


    Very similarly, one can evaluate the photon number correlation function in Eqn.\ref{nncorr}:
\begin{eqnarray}
    \langle {\cal T} n_{a}(\tau) n_a (0) \rangle - \langle n_a \rangle^{2}
     = \sum^{P}_{l=1} e^{-(E^{P}_{l}-E^P_0)\tau} | \langle P, l|n_{a} |P,G \rangle |^{2}
                                       \nonumber    \\
     \langle P, l|n_{a} |P,G \rangle = - \sum^{P}_{s=0}  A^{*P,l}_{s} A^{P,0}_{s} s,~~~ l\geq 1~~~~~~~~~
\label{nncorrnu}
\end{eqnarray}
    where $  \langle n_a \rangle= \sum^{P}_{s=0} | A^{P,0}_{s}|^{2} ( P- s )= \lambda^{2}_{a} $ and
    the Higgs mode $ E_{H}=E^{P}_{1}-E^P_0 $ with the spectral weight $ C_{H}= | \langle P, l=1|n_{a} |P,G \rangle |^{2} $.
    Very similarly, one can evaluate the atom correlation functions.

{\sl B. Relations between $ J-U(1)/Z_2 $ Dicke model and the $ U(1)/Z_2 $ model. }

  The energy levels in the lowest Landau level (LLL) shown in Fig.2 are identical
  in $ U(1) $ and $ J-U(1) $ models. This is because the ground state must be a totally symmetric state. In fact, every ground state
  in a given $ P= a^{\dagger} a + b^{\dagger} b $ sector must be a  totally symmetric state.
  It is the crossings of all these ground states at different $ P $ sectors which lead to all
  the energy levels in the LLL shown in the Fig.2.
  This explains why the diffusion constant $ D $ achieved by $ 1/J $ expansion in this paper is identical to that achieved by
  the $ 1/N $ expansion in Ref.\cite{berryphase}.  Because both  photon and total spin operators are
  also totally symmetric in the atom operators, then all the energy levels coupled to the ground state by the photon and total spin operators are also totally symmetric, so this also explains why we achieved the same
  single photon or atom correlation functions in the reduced Hilbert space in the $ J-U(1) $ Dicke model by $ 1/J $ expansion as those
  in the whole Hilbert space by the $ 1/N $ expansion in Ref.\cite{u1nature}.
  However, compared to the reduced Hilbert space in the $ J-U(1) $ Dicke model, there are many extra energy levels
  in the whole Hilbert space in the $ U(1) $ Dicke model, but they are not coupled to the ground state by the single
  photon or atom operators. Similar arguments apply to the more general $ U(1)/Z_2 $ model with the CRW term  in Eqn.1 and the $ J-U(1)/Z_2 $ model.

{\sl C. Comparisons with the Higgs mode and pseudo-Goldstone mode in one gap and two gaps superconductors }

     It is constructive to compare the Goldstone and Higgs mode of the atom-photo system studied in this report with those in  ( charge neutral ) superconductors ( so one can ignore the Anderson-Higgs mechanism for the sake of explaining physical concepts ).
     In a one gap superconductor, as explicitly demonstrated
     in the last reference in Ref.\cite{higgssuper}, when integrating out the fermions, the amplitude and phase of the
     paring order parameter $ \Psi= \Delta e^{i \theta} $ emerges as two {\sl independent } degree of freedoms, instead of being conjugate to each other.
     Its phase fluctuation in $ \theta $ leads to the Goldstone mode,
     while its amplitude fluctuation in $  \Delta $ leads to the Higgs mode.

   Now we consider the collective modes in a two gap superconductor such as $ Mg B_2 $ which has a $ \sigma $ band and a $ \pi $ band.
   Therefore it has two order parameters $ \Psi_{\sigma}=\Delta_{\sigma}e^{ i \theta_{\sigma} } $ and
   $ \Psi_{\pi} =\Delta_{\pi}e^{ i \theta_{\pi} }$. There are also fermionic degree of freedoms:
   $ \sigma $ electrons and $ \pi $ electrons.
   If ignoring the inter-band scattering $ V_{\sigma, \pi} $,
   the Hamiltonian has two independent $ U(1) $ symmetries: $ U(1)_{\sigma}\times U(1)_{\pi} $,
   the systems is just two copies of single band superconductor. So there are two independent Goldstone modes
   $ \theta_{\sigma}, \theta_{\pi} $ and also two independent Higgs modes $ \Delta_{\sigma}, \Delta_{\pi} $ for the two bands respectively.
   Now when considering the interband scattering term $ V_{\sigma, \pi} $, the symmetry of the Hamiltonian reduces from
   $ U(1)_{\sigma}\times U(1)_{\pi} $ to $ [U(1)_{\sigma}\times U(1)_{\pi}]_{D} $ where the $ D $ means the simultaneous rotation of the two
   order parameter phases. Then the two Goldstone modes couple to each other and
   split into one gapless Goldstone mode $ \theta_{+}=\theta_{\sigma}+ \theta_{\pi} $ plus a gapped pseudo-Goldstone mode
   $ \theta_{-}=\theta_{\sigma}- \theta_{\pi} $.
   The pseudo-Goldstone mode $ \theta_{-} $ is just the relative phase mode between the two order parameters
   whose gap is proportional to the strength of the interband scattering  $ V_{\sigma, \pi} $.
   The two Higgs modes $ \Delta_{\sigma}, \Delta_{\pi} $ will also couple to each other and  split into two new Higgs modes.
   In all, the two gaps superconductor has one gapless Goldstone mode and 3 gapped modes: one pseudo-Goldstone mode and two Higgs modes.

   A pseudo-Goldstone mode is always associated with an explicit symmetry breaking of a Hamiltonian, its gap is
   proportional to the strength of the explicit symmetry breaking.
   In contrast, a Higgs mode is the magnitude fluctuations of an order parameter. It is always associated
   with a spontaneous symmetry breaking in a ground state.
   The final physical meaning of a relative phase mode depends on the physical degree of freedoms of a system
   and its original relation to the order parameters of the system.
   As shown below Eqn.\ref{pmu10} in the main text, the conjugate pair $ ( \delta\rho_-, \theta_{-} ) $ fluctuation
   leads directly to the photon amplitude fluctuation mode, namely, the Higgs mode in Eqn.\ref{nncorr}.
   To some extent, the photon-atom system studied here is similar to one gap superconductor discussed in \cite{higgssuper}
   with the photon corresponding to the pairing order parameter, while the atoms corresponding to the fermions.
   When integrating out the atomic degree freedoms,
   the amplitude and phase of the photon order parameter emerges as two {\sl independent } degree of freedoms, instead of being conjugate to each other.
   Its phase fluctuation leads to the Goldstone mode, while its amplitude fluctuation leads to the Higgs mode.
   This fact was demonstrated by the $ 1/N $ expansion in \cite{berryphase} and also
   by Eqn.\ref{aacorr} and Eqn.\ref{nncorr} of this report by $ 1/J $ expansion.
   As shown in section D, a small counter-rotating wave $ g^{\prime} $ term in Eqn.\ref{u1}
   break sthe $ U(1) $ symmetry to a $ Z_2 $ symmetry, then the Goldstone mode at $ N=\infty $ will become a pseudo-Goldstone mode
   whose gap is proportional to the strength of the counter-rotating wave term.

{\sl D. The effects of the counter-rotating wave term at $ N=\infty $ and at a finite $ N $: }

  Now we consider the effects of the counter-rotating wave  (CRW ) terms in Eqn.\ref{u1}.
  Following the same procedures in the main text, we find that
  $  \lambda_a= \frac{g +  g^{\prime}}{ \omega_a } \sqrt{ \frac{j}{2} ( 1 - \mu^{2} ) }, \lambda_b= \sqrt{j(1-\mu) } $ where $
  \mu= \omega_a \omega_b/( g+ g^{\prime} )^{2} $, so the QCP is shifted to $  g+ g^{\prime}= g_{c}= \sqrt{ \omega_a \omega_b } $.
  The Hamiltonian to the order of $ 1/j $ is:
\begin{eqnarray}
    H_{U(1)/Z_2}  =  \frac{ D }{2} (\delta \rho_{+} - \alpha )^{2}
     + D_{-} [\delta \rho_{-} + \gamma  \delta \rho_{+} ]^{2}   \nonumber  \\
    +  4 \omega_{a} \lambda^{2}_a  \frac{ g }{ g + g^{\prime} } \sin^{2} \theta_{-}
     +  4 \omega_{a} \lambda^{2}_a  \frac{ g^{\prime} }{ g + g^{\prime} } \sin^{2} \theta_{+}
\label{u1z2h}
\end{eqnarray}
  where  $ D=  \frac{ 2 \omega_a (g+ g^{\prime} )^{2} }{  E^{2}_{H} N }  $ is the phase diffusion constant,
   $ D_{-}= E^{2}_{H}/16 \lambda^{2}_{a} \omega_a $ with $  E^{2}_{H}= ( \omega_a+\omega_b)^{2} + 4 ( g+ g^{\prime} ) ^2 \lambda^{2}_{a}/N $.
   The $ \gamma= \frac{ \omega^{2}_{a} }{ E^{2}_{H} } ( 1- \frac{ ( g + g^{\prime} )^{4}}{ \omega^{4}_{a} } ) $ is the coupling between the
   $ + $ and $ - $ sector.

   Eqn.\ref{u1z2h} can be rewritten as
\begin{equation}
    H_{U(1)/Z_2} = H_{U(1)}+ 2 \omega_{a} \lambda^{2}_a  \frac{ g^{\prime} }{ g + g^{\prime} } (1- \cos 2 \theta_{+} )
\label{u1z2hp}
\end{equation}
   where $ H_{U(1)} $ takes the same form as  Eqn.\ref{u1h} with the parameters
   corrected by $ g^{\prime} $. The last CRW term  breaks the $ U(1) $ symmetry to $ Z_2 $ symmetry
   $ \theta_{a/b} \rightarrow \theta_{a/b} + \pi $,
   $ \theta_{+} \rightarrow  \theta_{+} +  \pi, \theta_{-} \rightarrow  \theta_{-} $, so
   the $ \theta_{-} $ is neutral under the $ Z_2 $ transformation. In the thermodynamic limit $ N=\infty $,
   it leads to a small mass term for $ \theta_{+} $, so the Goldstone mode at $ N=\infty $ becomes a pseudo-Goldstone mode
   with a small gap $ \Delta_{PG}= \frac{4}{ E^{2}_{H} } \frac{ g^{\prime} }{ g + g^{\prime} } [ ( g+  g^{\prime})^{4}- g^{4}_{c} ] $.
   Obviously, this gap vanishes at the QCP $ g+ g^{\prime}= g_{c} $.
   In the following, we discuss its effects at a finite $ N $.

   If we ignore the CRW term, all the results achieved in the main text on the systems's energies Eqn.\ref{u1energy0}, the photon correlation function Eqn.\ref{aacorr} and  the photon number correlation function Eqn.\ref{nncorr} remain intact after making the corresponding changes in the
   parameters. Then for small $ g^{\prime}/g $, at a finite $ N $, we can can treat the CRW term by the perturbation theory.
   The calculations are straightforward and detailed in \cite{u1nature}. Here we only list the main results.
   Obviously, the high energy Higgs mode is in-sensitive to this CRW term, so we only need to focus on its effect on the low energy Goldstone mode.
   Then the sole dimensionless small parameter is  $ \delta= 2 \omega_{a} \lambda^{2}_a  \frac{ g^{\prime} }{ g + g^{\prime} }/ D $.
   (1)  For the Berry phase $ \alpha \neq 0 $, non-degenerate perturbation leads to the correction to the system's eigen-energy Eqn.\ref{u1energy0}
       at the second order $ \sim  \delta^{2} $. Note that although at $ \alpha = - 1/2 $, the energy is doubly degenerate
       with $ ( \delta \rho_{+}= m ,  \delta \rho_{+}= -m-1 ) $, but $ m $ and $ -m-1 $ carry opposite parities, so they will not be mixed by the CRW term.
       So the non-degenerate perturbation theory is valid. For the Berry phase $ \alpha = 0 $,
       because  the two degenerate states $ ( m , -m ), m > 0 $ carry the same parity,
       one need to use the degenerate perturbation theory to treat their splitting.
       The pair $ ( m , -m ) $ will split only at the $ m-$the order degenerate perturbation, so the splitting $ \Delta E  \sim  \delta^{m} $.
   (2) The normal photon correlation function Eqn.\ref{aacorr} receives a correction $ \sim \delta^{2} $ in both energy and spectral weight.
   Most importantly,  there appears also an anomalous photon correlation function $ \langle {\cal T} a(\tau) a (0) \rangle \sim \delta^{2} $.
   So the detection of a small anomalous photon correlation function by phase sensitive homodyne experiments \cite{exciton}.
   could be used to determine the strength of the CRW term.

   One can see that the corrections to all the physical quantities are at the second order $ \sim \delta^{2} $ or higher.
   From the $ N=2 $ qubits in the Fig.\ref{goldstone}a,
   one can see that  $ D \sim \omega_a/4 $,  $ 2 \lambda^{2}_{a} \sim 1 $ near the QCP, then when $ g^{\prime}/g < 1/3 $,
   the corrections due to the CRW term is suppressed compared to the finite size effects.
   Physically, at $ N=\infty $, any CRW term will transform the gapless Goldstone mode into a pseudo-Goldstone mode whose gap
   is proportional to the strength of the  CRW term. In contrast, at a finite $ N $,
   the quantum finite size effects already opened a gap to the Goldstone mode which is of the phase diffusion constant $ D \sim 1/N $.
   This gap make the Goldstone in a finite system $ N =2-5 $ quite robust against the CRW term if $ g^{\prime}/g < 1/3 $.

  In addition to  cold atoms inside an optical cavity or superconducting qubits \cite{qubitweak,qubitstrong} or quantum dots \cite{dots} inside a microcavity (Fig.\ref{mandel}a) discussed in the main text, there are also other promising experimental systems to realize the $ U(1) $ super-radiant phase.
  Most recently, the giant dipole moments of intersubband transitions in quantum wells have pushed the system into the ultrastrong
  light-matter coupling  regime in semiconductor heterostructures \cite{ultra}.
  Very recent experiments  \cite{elspin} achieved very strong coupling between an ensemble of $ s =1/2 $
  spins and photons in electronic spin ensembles coupled to superconducting cavities.
    The strong coupling regimes are also realized in ion Coulomb crystals in an
    optical cavity \cite{ion}. The CRW term could be easily suppressed to be small in these systems.
    Many new strong coupling light-matter systems with a small CRW term continue to emerge.



\begin{thebibliography}{99}



\bibitem{anderson} P.W. Anderson, Basic notions of condensed matter, 1983.

\bibitem{sachdev}  S. Sachdev, Quantum Phase transitions, 2012.

\bibitem{topo}
M. Z. Hasan and C. L. Kane, Colloquium: Topological insulators,  Rev. Mod. Phys. {\bf 82}, 3045 (2010);
X. L. Qi and S. C. Zhang, Topological insulators and superconductors, Rev. Mod. Phys. {\bf 83}, 1057 (2011).
" Exotic Phases of Frustrated Magnets " conference held at KITP  October 8-12, 2012.

\bibitem{higgs} F. Englert and R. Brout, Broken Symmetry and the Mass of Gauge Vector Mesons;
P.W. Higgs, Broken Symmetries and the Masses of Gauge Bosons, Phys. Rev. Lett. 13 (1964) 508;
G.S. Guralnik, C.R. Hagen, T.W.B. Kibble, Global Conservation Laws and Massless Particles, Phys. Rev. Lett. 13 (1964) 585.

\bibitem{higgsmag} Chubukov, A. V., Sachdev, S. and Ye, J. Theory of two-dimensional quantum Heisenberg antiferromagnets with a nearly critical ground state. Phys. Rev. B 49, 11919每11961 (1994)

\bibitem{jinwusuper}  Jinwu Ye and Longhua Jiang, Phys. Rev. Lett. 98, 236802 (2007); Jinwu Ye, Annals of Physics, 323 (2008), 580-630;
 Jinwu Ye, J. Low Temp Phys. 158(5), 882-900 (2010); 160(3), 71-111,(2010),
 Jinwu Ye, K.Y. Zhang, Yan Li, Yan Chen and W.P. Zhang,  Ann. Phys. 328 (2013) 103-138.


\bibitem{braggbog} M. Kozuma, {\sl et.al}, Phys. Rev. Lett. 82, 871
(1999); J. Stenger, {\sl et al}, Phys. Rev. Lett. 82, 4569 (1999);
D. M. Stamper-Kurn {\sl et al},
   Phys. Rev. Lett. 83, 2876 - 2879 (1999); J. Steinhauer, {\sl et.al}, Phys. Rev. Lett. 88,
   120407, (2002); S. B. Papp, {\sl et.al}, Phys. Rev. Lett. 101, 135301 (2008); P. T. Ernst, {\sl et al}, Nature Physics 6, 56 (2010 ).




\bibitem{higgscdw}  Sooryakumar, R. and Klein, M. Raman scattering by superconducting-gap excitations and their coupling to charge-density waves. Phys. Rev. Lett. 45, 660每662 (1980)

\bibitem{higgssuper}  Littlewood, P. and Varma, C. Gauge-invariant theory of the dynamical interaction of charge density waves and superconductivity. Phys. Rev. Lett. 47, 811每814 (1981), Amplitude collective modes in superconductors and their coupling to charge-density waves, Phys. Rev. B 26, 4883每4893 (1982).
    For a review, see Varma, C. Higgs Boson in superconductors. J. Low Temp. Phys. 126, 901每909 (2002); Ian J. R. Aitchison, Ping Ao, David J. Thouless, and X.-M. Zhu, Effective Lagrangians for BCS superconductors at T=0, Phys. Rev. B 51, 6531每6535 (1995).


\bibitem{higgssoft}  Podolsky, D. and Sachdev, S. Spectral functions of the Higgs mode near two-dimensional quantum critical points, Phys. Rev. B 86, 054508 (2012).

\bibitem{higgsmagvbs} R邦egg, C. et al. Quantum magnets under pressure: controlling elementary excitations in TlCuCl3. Phys. Rev. Lett. 100, 205701 (2008)



\bibitem{higgsbragg3d} Bissbort, U. {\sl et al.} Detecting the amplitude mode of strongly interacting lattice bosons by Bragg scattering. Phys. Rev. Lett. 106, 205303 (2011)

\bibitem{higgslattice2d} Manuel Endres, {\sl et.al }, The ＆Higgs＊ amplitude mode at the two-dimensional superfluid/Mott insulator transition,
                       Nature 487,454每458(26 July 2012).



\bibitem{higgslhc} ATLAS Collaboration, Observation of a new particle in the search for the Standard Model Higgs boson
with the ATLAS detector at the LHC, Physics Letters B 716 (2012) 1每29;
CMS Collaboration, Observation of a new boson at a mass of 125 GeV with the CMS experiment at
the LHC. Physics Letters B 716 (2012) 30每61.



\bibitem{orbitalt} A. T. Black, H. W. Chan and V. Vuletic, Observation of Collective Friction Forces due to Spatial Self-Organization of Atoms:
 From Rayleigh to Bragg Scattering, Phys. Rev. Lett. 91, 203001(2003).

\bibitem{orbital} K. Baumann, {\sl et.al}, Dicke quantum phase transition with a superfluid gas in an optical
cavity, Nature 464, 1301-1306 (2010);

\bibitem{fewboson} W. S. Bakr, {\sl et.al}, Probing the Superfluid每to每Mott Insulator Transition at the Single-Atom Level,
Science 30 July 2010: 547-550.

\bibitem{fewfermion}  F. Serwane, {\sl et.al},
Deterministic Preparation of a Tunable Few-Fermion System, Science 15 April 2011: 336-338.



\bibitem{qubitweak} A. Wallraff, {\sl et.al}, Strong coupling of a single photon to superconducting qubit using circuit quantum elctrodynamics,
Nature 431, 162-167 (2004)


\bibitem{qubitstrong} T. Niemczyk, {\sl et.al}, Circuit quantum electrodynamics in the ultrastrong-coupling regime, Nature Physics 6,772每776(2010).



\bibitem{dots} Reithmaiser, J. P, {\sl et.al}, Strong coupling in a single quantum dot-semi-conductor micro-cavity system,
Nature 432, 197-200 (2004).
Yoshie, T. {\sl et al}, Vacuum Rabi splitting with a single quantum dot in a photonic crystal nanocavity, Nature 432, 200-203 (2004).
K. Hennessy, A. Badolato, M. Winger, D. Gerace, M. Atat邦re, {\sl et al}, Quantum nature of a strongly coupled single quantum dot每cavity system,
Nature 445, 896-899 (22 February 2007).



\bibitem{dicke} R. H. Dicke,   Phys. Rev. 93, 99 (1954).

\bibitem{dicke1} K. Hepp and E. H. Lieb, Anns. Phys. ( N. Y. ), 76,
360 (1973); Y. K. Wang and F. T. Hioe,  Phys. Rev. A, 7, 831 (1973).

\bibitem{popov} V. N. Popov and S. A. Fedotov,  Soviet Physics JETP, 67, 535 (1988); V. N. Popov and V. S. Yarunin, Collective Effects in Quantum
Statistics of Radiation and Matter (Kluwer Academic,
Dordrecht,1988).


\bibitem{bethe} The $ U(1) $ Dicke ( Tavis-Cummings ) model is
integrable at any finite $ N $, so, in the " face " value, the
system's eigen-energy spectra  could be "exactly" solvable by Bethe
Ansatz like methods. For example, see N.M. Bogoliubov, R.K.
Bullough, and J. Timonen, Exact solution of generalized
Tavis-Cummings models in quantum optics, J. Phys. A: Math. Gen. 29
6305 (1996). However, so far, the Bethe Ansatz like solutions stay
at very "formal" level from which it is even not able to get the
system's eigen-energy levels Eqn.\ref{u1energy0} analytically, let alone to extract any
underlying physics explored in this paper. Furthermore, it is known that the Bethe Ansatz
method is not able to get any  dynamic correlation functions.

\bibitem{staircase} V. Buzek, M. Orszag and M. Roko,  Phys. Rev. Lett. 94, 163601 (2005).

\bibitem{berryphase}  Jinwu Ye  and  CunLin Zhang, Super-radiance, Photon condensation  and its phase diffusion,
        Phys. Rev. A 84, 023840 (2011).

\bibitem{chaos} C. Emary and T. Brandes,
 Phys. Rev. Lett. 90, 044101 (2003); Phys. Rev. E 67, 066203
(2003). N. Lambert, C. Emary, and T. Brandes, Phys. Rev. Lett. 92,
073602 (2004).
In a recent unpublished work, using the $ 1/J $ expansion and the ED, the authors studied
the $ Z_2 $ Dicke model at a finite $ N $. Due to the very different
symmetries, the $ U(1) $ and $ Z_2 $ Dicke models show completely
different properties.



\bibitem{gprime1} F. Dimer, B. Estienne, A. S. Parkins, and H. J. Carmichael, Phys. Rev. A, 75, 013804, 2007

\bibitem{gprime2} M. J. Bhaseen, J. Mayoh, B. D. Simons1, and J. Keeling, Dynamics of nonequilibrium Dicke models, Phys. Rev. A, 85, 013817 (2012).

\bibitem{u1nature} Yu Yi-Xiang  {\sl et.al}, unpublished.

\bibitem{nobel1} J. C. Bergquist, {\sl et.al},
Observation of Quantum Jumps in a Single Atom, Phys. Rev. Lett. 57, 1699每1702 (1986);
C. Monroe, {\sl et.al},
Demonstration of a Fundamental Quantum Logic Gate, Phys. Rev. Lett. 75, 4714每4717 (1995)

\bibitem{nobel2} M. Brune,{\sl et.al},
Quantum Rabi Oscillation: A Direct Test of Field Quantization in a Cavity, Phys. Rev. Lett. 76, 1800每1803 (1996);
M. Brune, {\sl et.al},
Observing the Progressive Decoherence of the ※Meter§ in a Quantum Measurement, Phys. Rev. Lett. 77, 4887每4890 (1996);
M. Brune, {\sl et.al}, Manipulation of photons in a cavity by dispersive atom-field coupling: Quantum-nondemolition measurements and generation of ＆＆Schrodinger cat＊＊ states, Phys. Rev. A 45, 5193每5214 (1992).

\bibitem{exciton} Jinwu Ye, T. Shi and Longhua Jiang, Phys. Rev. Lett. 103, 177401 (2009);
  T. Shi,  Longhua Jiang and Jinwu Ye, Phys. Rev. B 81, 235402 (2010);
  Jinwu Ye, Fadi Sun, Yi-Xiang Yu  and Wuming Liu,  Ann. Phys. 329, 51每72 (2013).



\bibitem{ultra} G. Gunter, {\sl et.al}, Sub-cycle switch-on of ultrastrong light-matter
interaction, NATURE, Vol 458, 178,  12 March 2009.
Aji A. Anappara1, {\sl et.al},
Signatures of the ultrastrong light-matter coupling regime, Phys. Rev. B 79, 201303(R) (2009).

\bibitem{elspin}  D. I. Schuster, {\sl et.al} High-Cooperativity Coupling of Electron-Spin Ensembles to Superconducting
   Cavities, Phys. Rev. Lett. 105, 140501 (2010);
   Y. Kub, {\sl et.al}, Strong Coupling of a Spin Ensemble to a Superconducting
   Resonator, Phys. Rev. Lett. 105, 140502 (2010).

\bibitem{ion} Peter F. Herskind, {\sl et.al}, Realization of collective
strong coupling with ion Coulomb crystals in an optical cavity,
NATURE PHYSICS, VOL 5, 494, JULY 2009.

\end{thebibliography}
\end{document}